\documentclass[12pt]{article}
\usepackage{pic03}
\usepackage{hyperref}
\usepackage{url}
\usepackage{graphicx}

\newcommand{\snoccfluxshort}{1.76^{+0.06}_{-0.05}\mbox{(stat.)}^{+0.09}_{-0.09}~\mbox{(syst.)}} 
\newcommand{\snoesfluxshort}{2.39^{+0.24}_{-0.23}\mbox{(stat.)}^{+0.12}_{-0.12}~\mbox{(syst.)}} 
\newcommand{\snoncfluxshort}{5.09^{+0.44}_{-0.43}\mbox{(stat.)}^{+0.46}_{-0.43}~\mbox{(syst.)}} 
 
\newcommand{\snomutauflux}{3.41^{+0.45}_{-0.45}\mbox{(stat.)}^{+0.48}_{-0.45}~\mbox{(syst.)}} 
\newcommand{\snoeflux}{1.76^{+0.05}_{-0.05}\mbox{(stat.)}^{+0.09}_{-0.09}~\mbox{(syst.)}}

\newcommand{\ssmflux}{5.05^{+1.01}_{-0.81}}

\newcommand{\ns}{$^{16}$N}

\begin{document}

\title{\bf LATEST NEWS FROM SNO}
\author{K. Graham    for the SNO Collaboration   \\
{\em Department of Physics, Queen's University, Kingston, ON, Canada K7L
3N6}}
\maketitle

\vspace{4.5cm}

\baselineskip=14.5pt
\begin{abstract}
From the initial pure D$_2$O phase, the SNO experiment has measured the
flux of $^8$B solar neutrinos via 
CC, NC, and ES interactions.  The NC flux
measurement, $\phi^{\mbox{\scriptsize SNO}}_{\mbox{\scriptsize NC}} =
5.09^{+0.64}_{-0.61} \times 10^{6}\mbox{cm}^{-2} \mbox{s}^{-1}$,
represents the total flux of active neutrinos and is consistent with
solar model expectations.  When combined with the CC,
 $\phi^{\mbox{\scriptsize SNO}}_{\mbox{\scriptsize CC}} = 
1.76^{+0.11}_{-0.10}\times 10^{6} \mbox{cm}^{-2} \mbox{s}^{-1}$,
and ES, $\phi^{\mbox{\scriptsize SNO}}_{\mbox{\scriptsize ES}} =
2.39^{+0.27}_{-0.26}
\times 10^{6} \mbox{cm}^{-2} \mbox{s}^{-1}$ results, SNO demonstrates
the existence of neutrino flavour transformations at the 5.3$\sigma$ level.

The second salt phase of SNO operation is nearing completion.  The
detector calibration for this phase has been carried out and
systematic uncertainties have been evaluated in preparation for the first salt
publication.  As demonstrated from MC simulation studies, 1 year of
salt data livetime will produce CC and ES results with similar precision as
the pure D$_2$O phase and a substantially improved NC measurement.  In
addition, precision flux measurements with no energy constraint will
be possible.
\end{abstract}
\newpage

\baselineskip=17pt

\section{Introduction}
The Sudbury Neutrino Observatory (SNO)~\cite{NIM}, a 1000 T D$_2$O Cerenkov detector
located 2 km underground (6010 m water equivalent) in Sudbury, Canada,
is sensitive to $^8$B solar neutrinos via the interactions
\begin{eqnarray}
  &&\nu_{\mbox{e}}  +  \mbox{d} \;\;\rightarrow \;\mbox{p} + \mbox{p} +
  \mbox{e}^{-}  \;\;(\mbox{CC}) \nonumber \\
  &&\nu_{\mbox{x}}  +  \mbox{d} \;\;\rightarrow \;\mbox{p} + \mbox{n} +
  \nu_{\mbox{x}}  \;\;(\mbox{NC}) \nonumber \\  
  &&\nu_{\mbox{x}}  +  \mbox{e}^- \rightarrow\; \nu_{\mbox{x}} +
  \mbox{e}^-  
\;\;\;\;\;\;\;(\mbox{ES})
\end{eqnarray}
where  $\nu_{\mbox{x}}$ represents all active neutrino flavours.

Whereas the charged current (CC) interaction is only sensitive to
$\nu_{\mbox{e}}$ neutrinos, the neutral current (NC) interaction is
equally sensitive to all active neutrino flavours.  This sensitivity to
non-$\nu_{\mbox{e}}$ neutrinos, combined with the precise
$\nu_{\mbox{e}}$ flux measurement from the CC channel, gives SNO the
unique ability to test the hypothesis that a portion of the
$\nu_{\mbox{e}}$'s produced in the sun undergo flavour change while in
transit to earth.  The elastic scattering (ES) interaction does have some sensitivity to non-$\nu_{\mbox{e}}$ neutrinos but for SNO is a low statistics channel.

Since the standard solar model predicts a negligible flux of
non-$\nu_{\mbox{e}}$ neutrinos, a non-zero difference between the
total flux of active neutrinos from the NC channel and the flux of
electron neutrinos from the CC channel would provide a clear
indication of neutrino flavour transformation.

SNO has nearly completed the second of the first two phases of data taking.
Results from the initial `pure D$_2$O' phase, in which the central target consisted
solely of purified heavy water, are described in the next section of
this document.  
 The final section provides  an update of the status
of the second `salt phase' in which the characteristic response of NC
events has been altered by the addition of NaCl to the D$_2$O.

\section{Pure D$_2$O Phase Results}
\subsection{SNO Detector, Calibration, and Backgrounds}

The SNO Detector consists of 9456 photo-multiplier tube (PMT) and light
concentrator units arrayed on a geodesic support structure.  This array
surrounds a 1700 T spherical shell of light water, which in turn
encloses the spherical acrylic vessel containing the 1000 T of heavy water.

Electrons produced in the D$_2$O emit a cone of Cerenkov light which is
detected by the PMT's.  Neutrons produced in the detector can be
captured by a deuteron which subsequently emits a 6.25 MeV photon.
Photons interact with electrons in the D$_2$O via Compton scattering and
these electrons are detected as described above.

The event energy is approximately linearly related to the number of
PMT's registering light.  Vertex positions and event directions are
reconstructed via fitting algorithms which use the relative PMT
positions and firing times to reconstruct the Cerenkov cone.

In order to understand the detector response and systematic
uncertainties, a series of calibration sources are placed at various
positions within the detector volume.  Analyses of the calibration data
and comparisons with Monte Carlo simulated events are then carried out.

The optical properties of the detector are evaluated by transmitting
laser light at wavelengths between 337-620 nm into the D$_2$O volume via
optical fibers affixed to a glass diffusing ball.  The response of the PMT
array, for various positions of the diffusing ball, is evaluated and
the optical properties of the detector measured.

In order to set the absolute energy scale and to evaluate energy scale
and resolution uncertainties, an $^{16}$N source\cite{n16}, which emits 6.13 MeV photons,
is utilized.  These studies are supplemented by $^8$Li source~\cite{li8} runs in
which $\beta$'s with an end point near 14 MeV are produced and 
$^3$H(p,$\gamma$)$^4$He(`pT') source~\cite{poon} runs in which 19.8 MeV photons
are emitted.  In addition to the energy response, these sources are
also used to determine the reconstructed event vertex and event
direction detector response and uncertainties.

The NC flux measurement requires an understanding of the detector
neutron capture efficiency.  A $^{252}$Cf source is introduced at
various positions within the detector to produce a known quantity of
neutrons and analysis of this data provides a measurement of the
capture efficiency and uncertainties. 

Evaluations of backgrounds in the data set are carried out through a
variety of ex-situ and in-situ analyses.  Instrumental backgrounds are
evaluated through an examination of calibration data and a series
of selection cuts adopted to suppress them.  
Ex-situ measurements using ion exchange and membrane degassing
techniques coupled with in-situ analyses of the SNO low-energy signal
region are employed to estimate the level of radioactive thorium
and uranium backgrounds.  These backgrounds give rise to 
photons that contaminate the low-energy region of the SNO spectrum
and, more crucially, can cause photo-disintegration of the deuterons
inducing a neutron background signal indistinguishable from NC events.
\begin{table}
\begin{tabular}{llllll}
Source        &  CC Uncert. \%          & NC Uncert. \%        	&
$\phi_{\mu\tau}$ Uncert. \%   \\
\hline
Energy scale   		& -4.2,+4.3 &-6.2,+6.1 		&-10.4,+10.3  \\ 
Energy resolution   	& -0.9,+0.0 &-0.0,+4.4 		&-0.0,+6.8  \\ 
Vertex accuracy 				& -2.8,+2.9 &  $\pm 1.8$ 	&  $\pm 1.4$    \\ 

D$_2$O Cerenkov   	& -0.1,+0.2 &-2.6,+1.2 		&-3.7,+1.7  \\

PMT Cerenkov   		& $\pm 0.1$ &  -2.1,+1.6 	&-3.0,+2.2  \\ 
Neutron capture 				& $\pm 0.0$ &  -4.0,+3.6 	&-5.8,+5.2  \\ 
\hline
Total uncertainty 		& -5.2,+5.2 &  -8.5,+9.1 	&  -13.2,+14.1 \\ \hline 
\end{tabular}
\caption{\label{tab:sys}The primary systematic uncertainties on the
extracted charged current (CC) and neutral current (NC) neutrino fluxes are presented.  In addition, the uncertainties on the measured non-electron neutrino flux ($\phi_{\mu\tau}$) are given.  The total uncertainty for the elastic scattering flux (ES - not shown) is -4.8,+5.0 percent.}
\end{table}

The combined systematic uncertainties from these studies are presented in
Table \ref{tab:sys}.

\subsection{Pure D$_2$O Phase Results}

After applying data cuts to reduce backgrounds, including a lower
kinetic energy threshold cut of T$\ge5$ MeV, and event vertex radius
cut R$\le550$ cm, 2928 events are selected from the 306.4 days of
live time collected between Nov. 2, 1999 and May 28, 2001.

The reconstructed energy, angle between event direction and direction
to the sun, and vertex radius spectra from this sample are
simultaneously fit using an extended maximum likelihood technique to
estimate the relative fractions of CC, NC, and ES events in the
sample.  

\begin{figure}[htbp]
  \centerline{\hbox{ \hspace{0.2cm}
    \includegraphics[width=7.5cm]{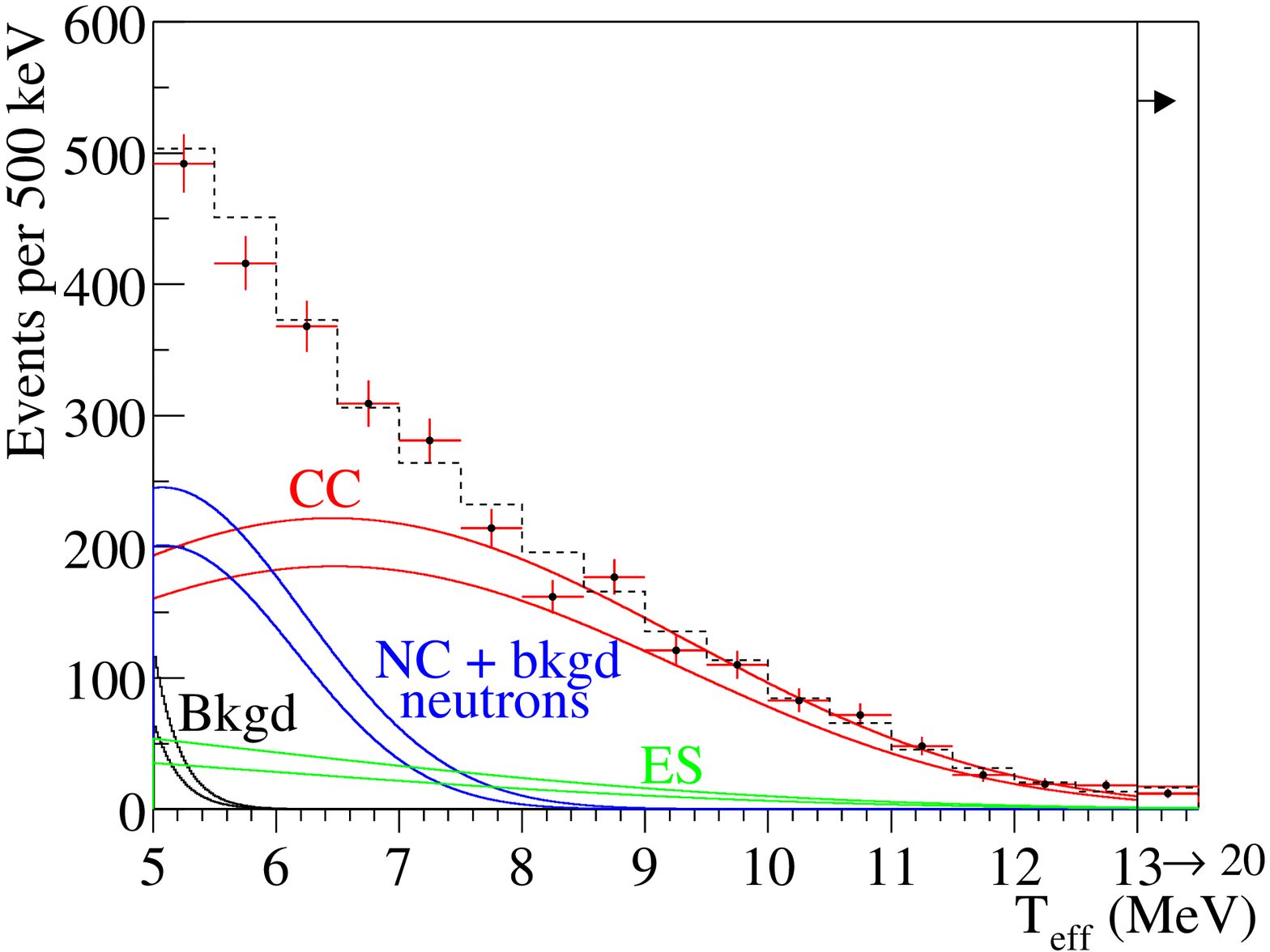}
    \hspace{0.15cm}
    \includegraphics[width=7.5cm]{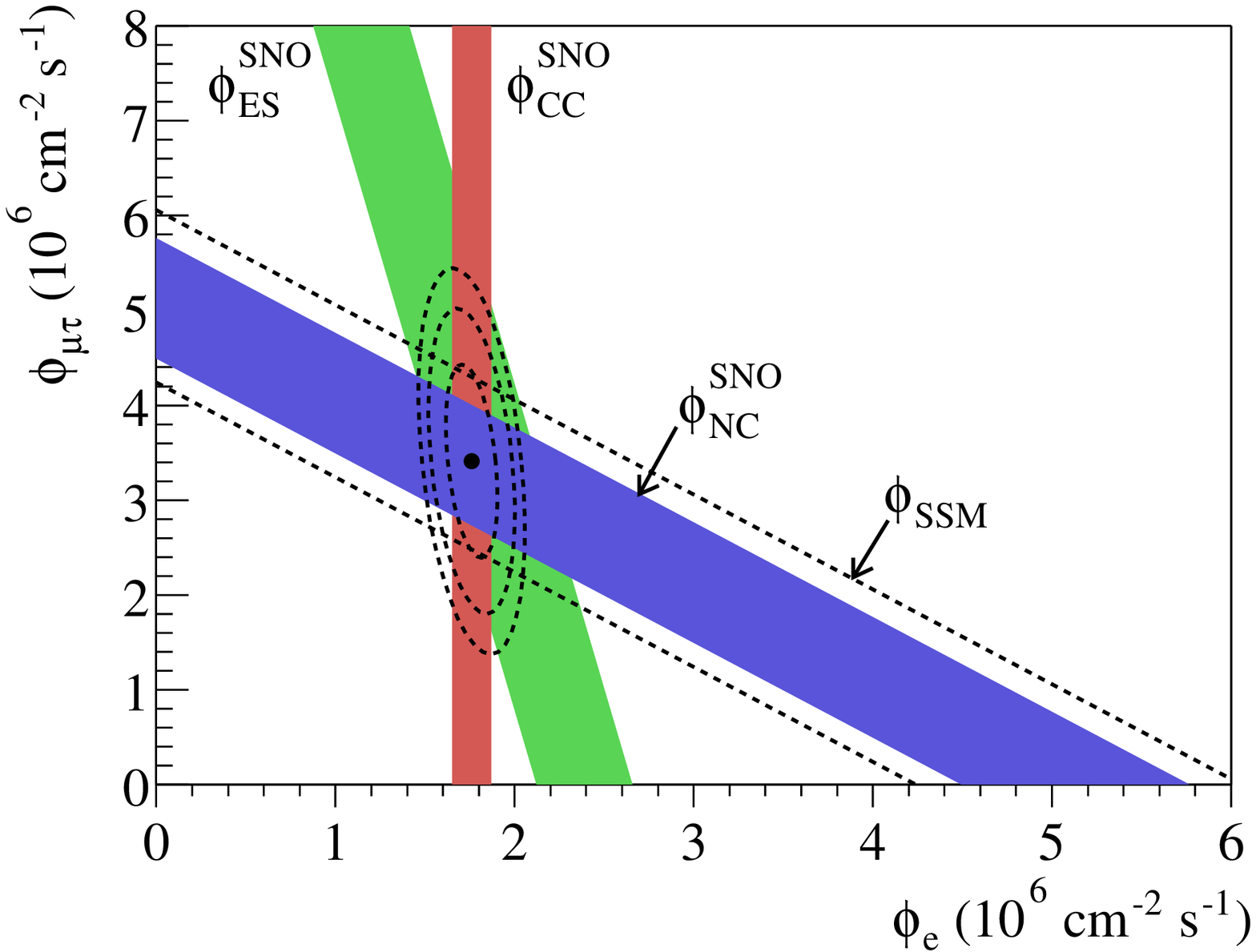}
    }
  }
 \caption{\it Shown are the pure D$_2$O phase
   kinetic energy spectrum of neutrino events (points) and fit
   distribution (solid) (left) and confidence level regions in the
$\phi_{\mu\tau}$-$\phi_{\mbox{\scriptsize e}}$ plane (right).}
    \label{fig:fitresults} 
\end{figure}

The fit results, assuming a standard $^8$B spectral shape\cite{ortiz},
are $1967.7^{+61.9}_{-60.9}$ CC events, $263.6^{+26.4}_{-25.6}$ ES events, and
$576.5^{+49.5}_{-48.9}$  NC events; only statistical
uncertainties are given. Figure \ref{fig:fitresults} shows the energy
spectrum of the data overlayed by the fit distribution and the
relative contributions from CC, ES, NC, and backgrounds.  

The measured fluxes of $^8$B neutrinos
from each of the three channels, including systematic uncertainties,
are ~\cite{ccprl}\cite{ncprl}
\begin{eqnarray}
\;\;\phi^{\mbox{\scriptsize SNO}}_{\mbox{\scriptsize CC}} & = & \snoccfluxshort \times 10^{6}
\mbox{cm}^{-2} \mbox{s}^{-1}\\
\;\;\phi^{\mbox{\scriptsize SNO}}_{\mbox{\scriptsize ES}} & = & \snoesfluxshort \times 10^{6}
\mbox{cm}^{-2} \mbox{s}^{-1}\\
 \label{eq:ncresult} \;\;\phi^{\mbox{\scriptsize SNO}}_{\mbox{\scriptsize NC}} & = & \snoncfluxshort \times 10^{6}
\mbox{cm}^{-2} \mbox{s}^{-1}. 
\end{eqnarray}

Transforming these numbers into estimates of the flux of
$\nu_{\mbox{e}}$ and the flux of $\nu_{\mu\tau}$ (assuming only three
active neutrino flavours) gives
\begin{eqnarray}
\phi_{e}\;\; & = & \snoeflux \times 10^{6}
\mbox{cm}^{-2} \mbox{s}^{-1}\\
\;\;\phi_{\mu\tau} & = & \snomutauflux \times 10^{6}
\mbox{cm}^{-2} \mbox{s}^{-1}.
\end{eqnarray}

The measured non-zero flux of $\nu_{\mu\tau}$ neutrinos indicates, at
a more than 5$\sigma$ level, that $\nu_{\mbox{e}}$ produced in the sun
undergo flavour transformation.  Figure \ref{fig:fitresults} indicates
this result graphically, where it may be seen that the $\phi_{e}$
measurement comes almost entirely from the CC result whereas the
$\phi_{\mu\tau}$ estimate is primarily given by the NC result. 
The ES result does provide additional information for both fluxes and
is in good agreement with the CC and NC values.

The NC measured total flux of active $^8$B neutrinos 
given in Equation \ref{eq:ncresult} is consistent with the standard solar model
theoretical prediction~\cite{BPB} of $\phi_{\texttt{SSM}} = \ssmflux.$
However, this result requires constraining the shape of the $^8$B
energy spectrum.  Relaxing this constraint produces the less precise result
\begin{equation}
\phi_{\texttt{SNO}} = 6.42^{+1.57}_{-1.57}
(stat.)^{+0.55}_{-0.58}(sys.).
\end{equation}

\begin{figure}[htbp]
  \centerline{\hbox{ \hspace{0.2cm}
    \includegraphics[width=7.cm]{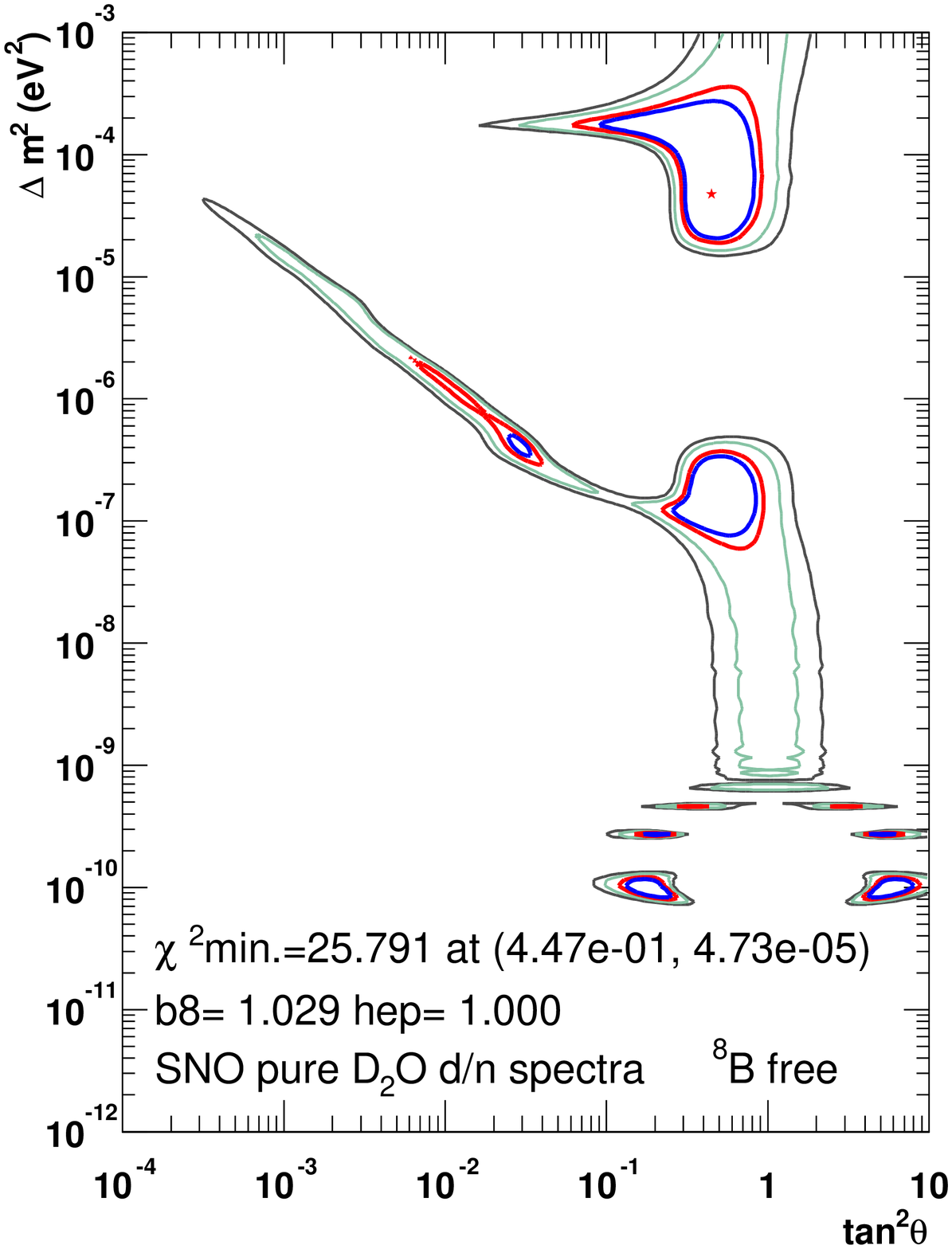}
    \hspace{0.15cm}
    \includegraphics[width=7.cm]{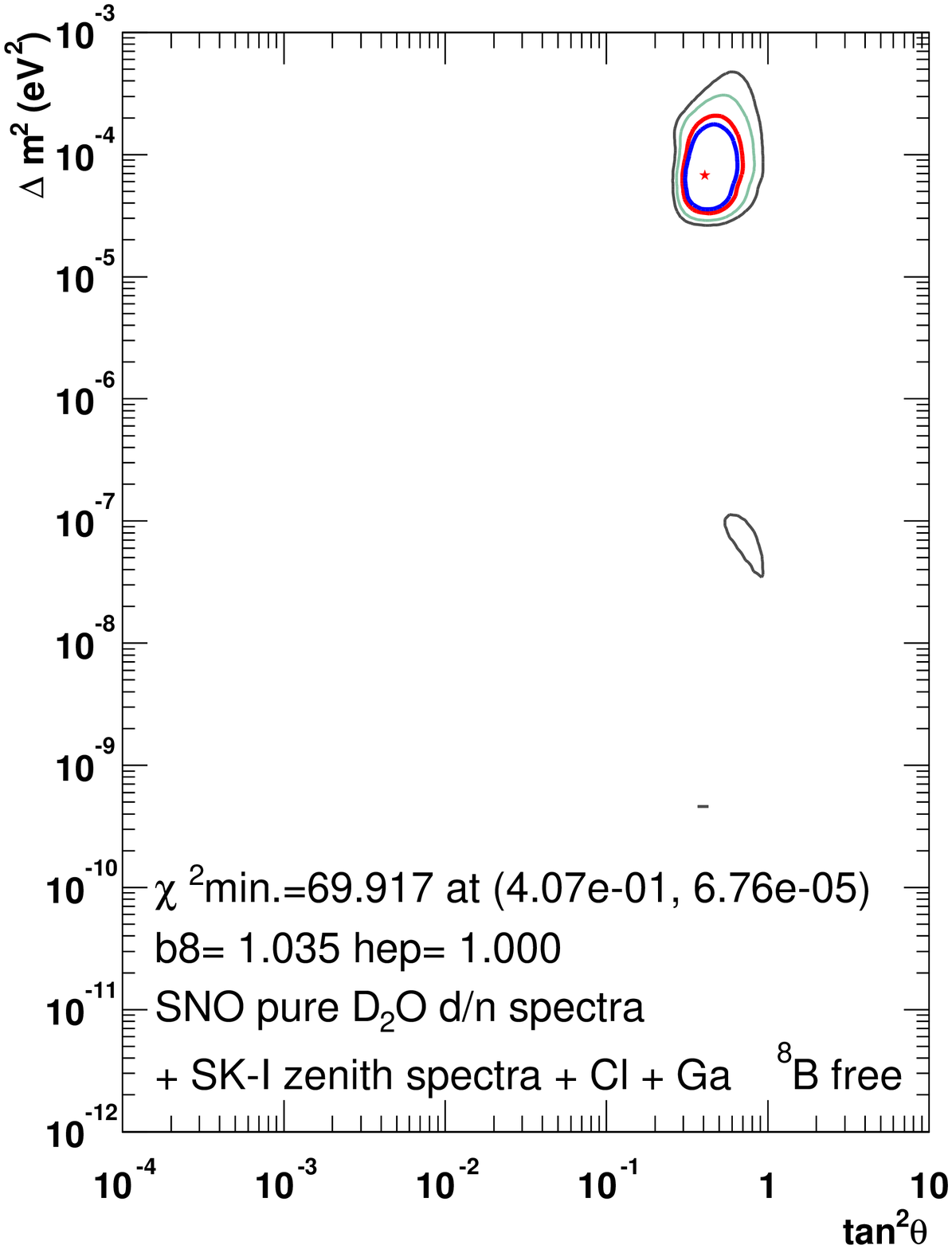}
    }
  }
 \caption{\it Shown are
MSW C.L. region from SNO only pure D$_2$O data (left) and 
for combined solar neutrino data (right).}
    \label{fig:msw1} 
\end{figure}

One of the leading theoretical candidates to explain neutrino flavour
transformations is the MSW model~\cite{msw}. 
SNO data has been evaluated in the context of this model producing the
confidence level regions in the $\Delta$m$^2$-$\tan^2{\theta}$ plane given
in Figure \ref{fig:msw1}~\cite{dnprl}.  As can be seen, a large portion of MSW
space is rejected at the 90\% C.L. by SNO data.  

Combining the SNO measurements with results from Cl \cite{cl}, Ga \cite{gallex,sage,gno},
and the Super-Kamiokande \cite{SK} experiments produces the
confidence level contours given in Figure \ref{fig:msw1}.  As shown,
current experimental evidence from solar neutrino experiments
discounts all but LMA and LOW region solutions at better than a 99\%
C.L..  The relative minima in these two regions strongly favours the LMA
solution.

\section{Salt Phase Update}

In May 2001, approximately 2 tonnes of NaCl was dissolved in the D$_2$O
volume of the SNO detector.  The addition of the salt enhanced the
neutron capture efficiency and the characteristic response to NC
events.  The first salt publication will focus on an initial data set
corresponding to roughly 250 live days of running.

\subsection{Detector Calibration and Response}

As with the pure D$_2$O phase, the detector response must be
calibrated and systematic uncertainties evaluated for the salt phase.
Three examples highlighting the current status of calibrations in the
salt phase are given here.

The energy scale calibration is carried out via a
combination of laser and \ns~ scans as outlined above.  Full laser
optical scans are taken at approximately 3-4 month intervals.  \ns~
stability runs with the source positioned at the centre of the
detector are taken every few weeks while full two-plane scans with
the source located at many source positions throughout the D$_2$O
volume are taken every few months.

\begin{figure}[htbp]
  \centerline{\hbox{ \hspace{0.2cm}
    \includegraphics[width=7.5cm]{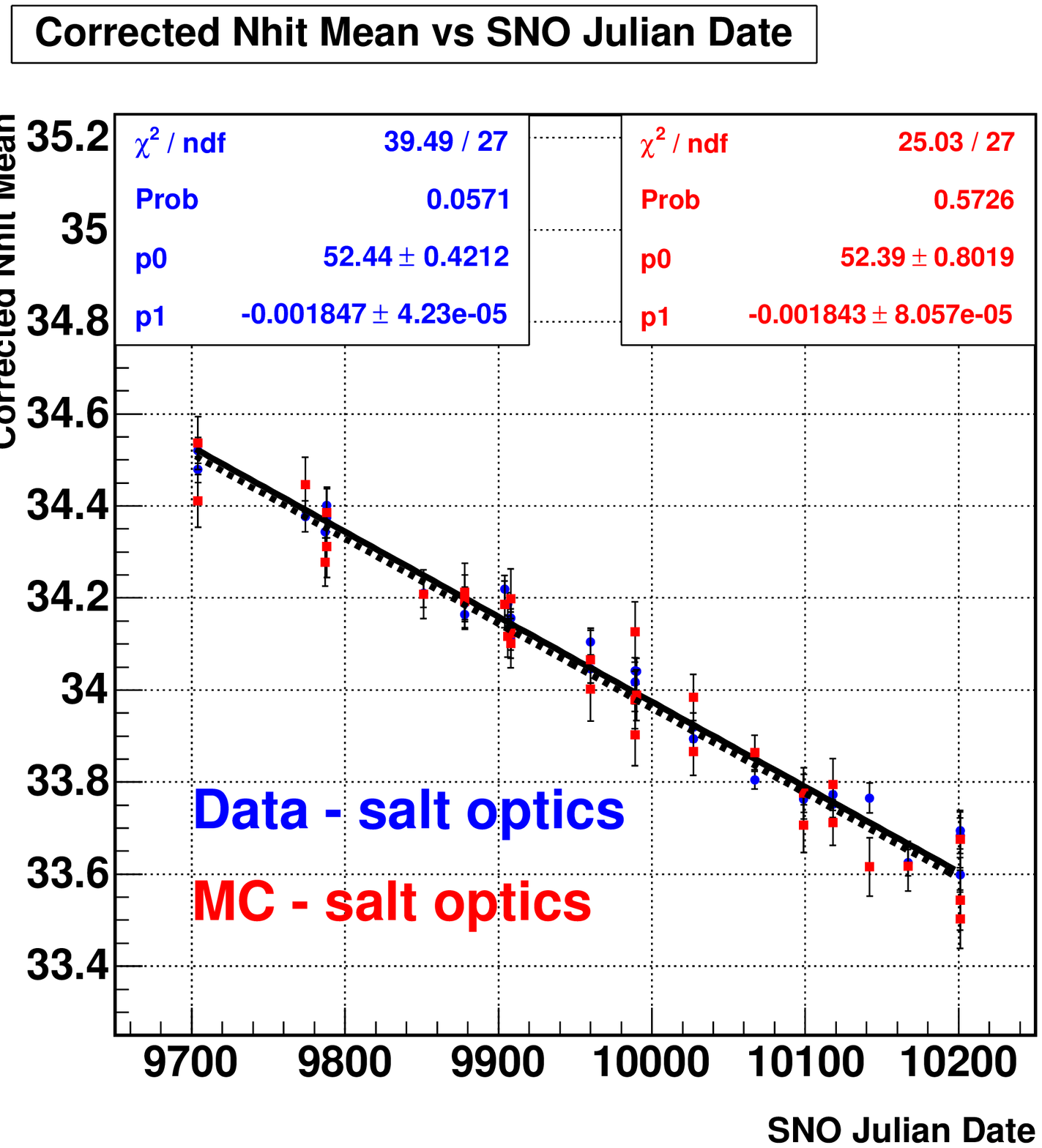}
    \hspace{0.15cm}
    \includegraphics[width=7.5cm]{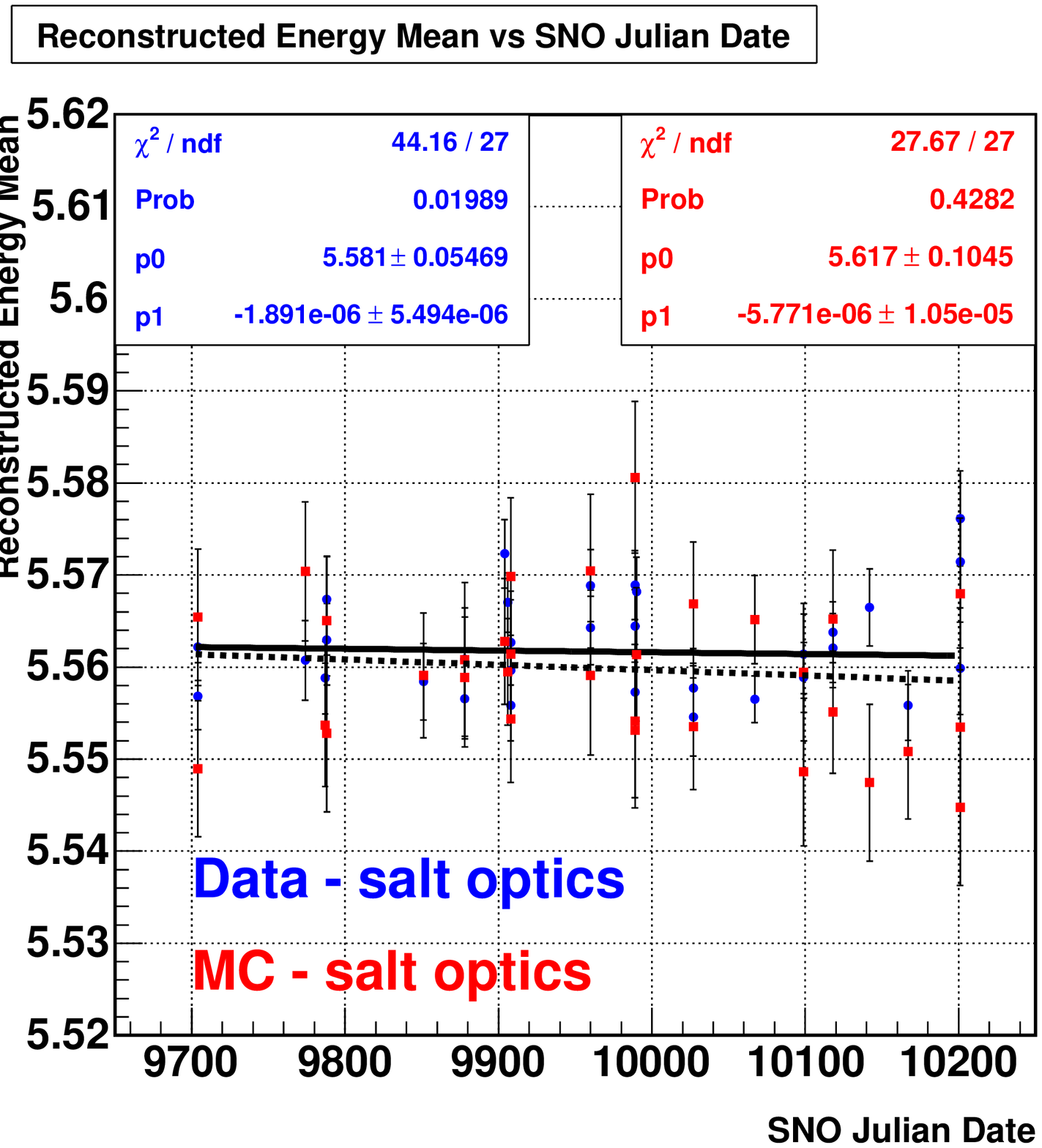}
    }
  }
 \caption{\it
   Mean effective number of PMT's triggering per \ns~ event (left)
   for centrally positioned data (blue) and MC
   (red) stability runs and mean reconstructed energy (right)
   distributions as a function of run date are shown.
    \label{fig:response} }
\end{figure}

Early in the salt phase it was discovered that the energy response
of the detector was changing with time.  Figure \ref{fig:response}
shows the mean effective number of PMT's (Neff) triggering per \ns~ event as a
function of run date for \ns~ stability runs.  As can be seen, the
response changes by approximately $\sim2\%$/year.

Subsequent evaluation of the laser data indicates that the measured
D$_2$O attenuations lengths were changing in a corresponding fashion.
MC simulated \ns~ events have been generated assuming time varying
attenuations and the mean effective number of PMT's for each MC run
are shown in Figure \ref{fig:response}.  As can be seen, the change in
response predicted by the MC for the measured changing attenuations
agrees well with the measured change in the \ns~ data.

MC simulated electrons at a series energies are generated using the
laser scan derived optical constants and a mapping between Neff and
energy derived.  After applying all corrections and this mapping
function to the \ns~ events, the mean energy response shown in Figure
\ref{fig:response} is produced.  Systematic uncertainties associated
with the stability of detector response in time are evaluated by
examining the variations in and differences between the data and MC
\ns~ energy distributions.

Systematic uncertainties associated with spatial
variations and asymmetries in the detector are evaluated by examining
\ns~ data and MC runs taken at various positions throughout the D$_2$O
volume.  Including additional uncertainties associated with source
modeling, rate dependence, hardware stability, timing calibration,
and cross-talk the total energy scale systematic uncertainty is
roughly percent level.

The primary method for distinguishing ES events from other signal and
background events is to exploit the fact that the reconstructed event
directions for such events will be correlated with the direction to the
sun.  Hence a good understanding of the detector angular response is
required.

Data and MC \ns~ events are used to evaluate the systematic
uncertainties on the angular response.  What is important here is the
agreement between data and MC spatially and temporally.  

For these events, the `generated' event direction is taken to be the
direction from the \ns~ source position to the reconstructed event
vertex.  To ensure a reliable estimate for the `generated' direction,
a cut is placed requiring the reconstructed vertex of the electron to
be at least 150 cm away from the \ns~ source
position.  

\begin{figure}[htbp]
  \centerline{\hbox{ \hspace{0.2cm}
    \includegraphics[width=7.5cm]{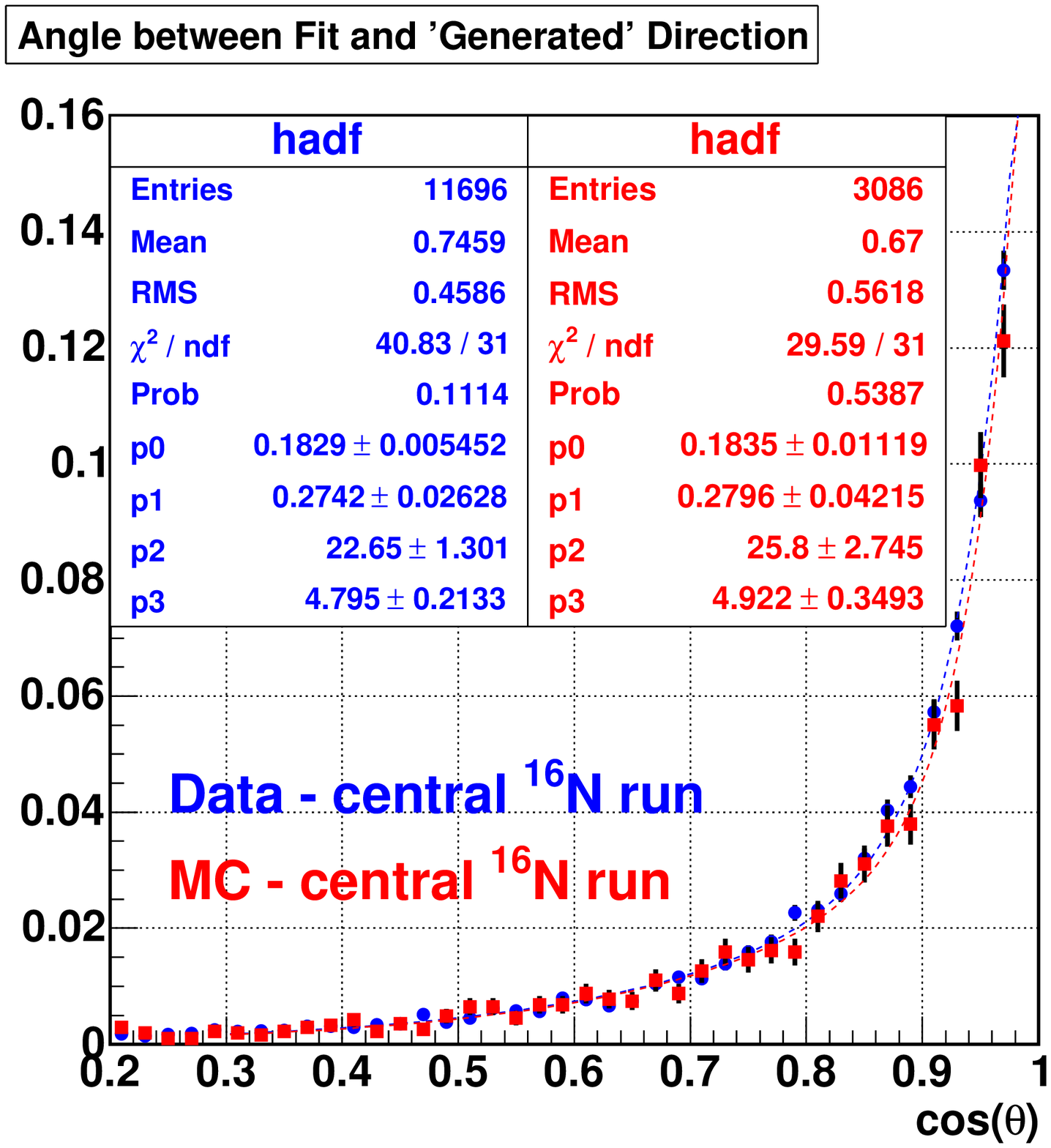}
    \hspace{0.15cm}
    \includegraphics[width=7.5cm]{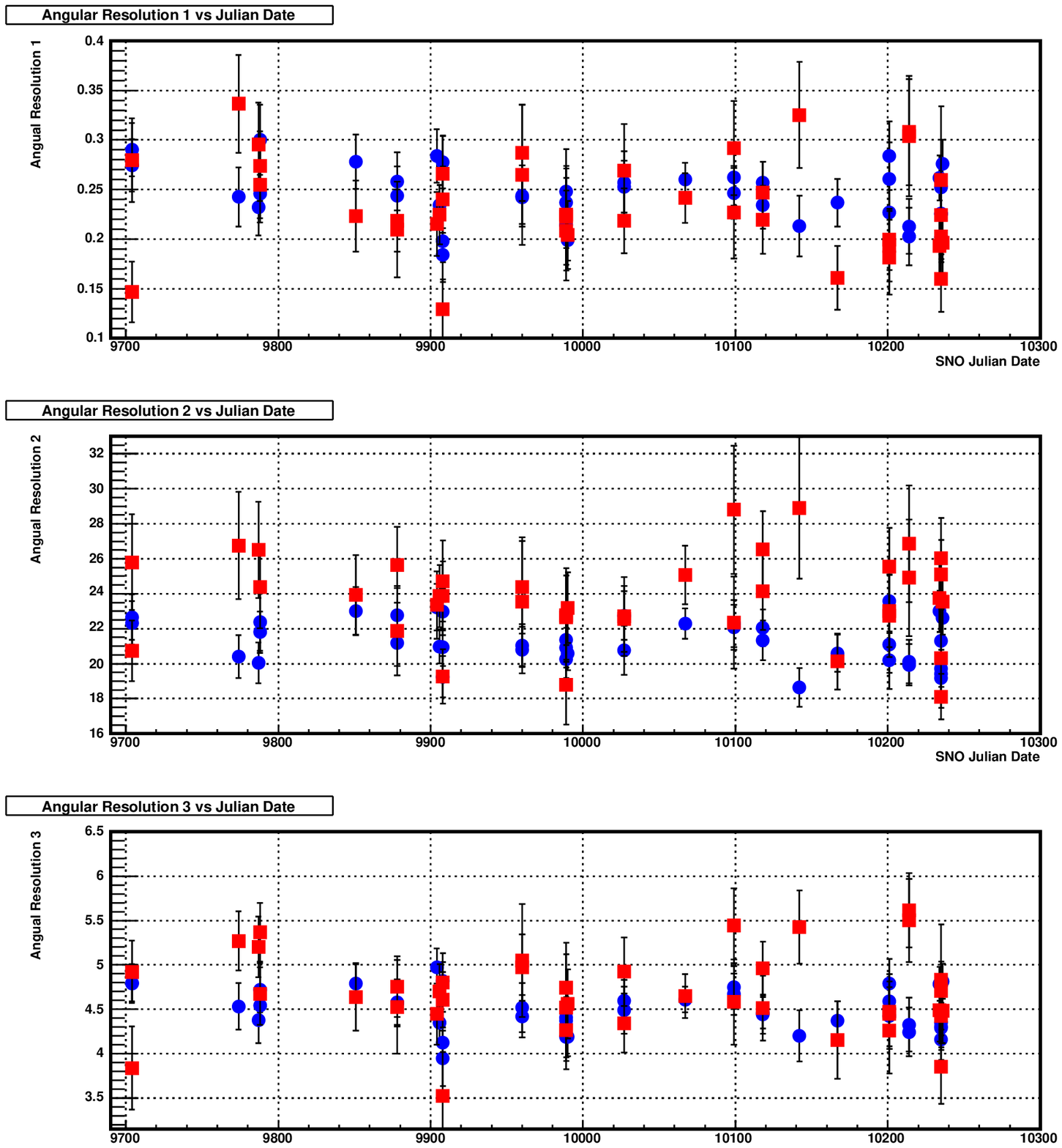}
    }
  }
 \caption{\it
Shown are the distribution of the cosine of the
angle between the fit and `generated' direction for data (blue) and MC
(red) \ns~events for a run taken at the centre of the detector (left) and fit
parameters for data (blue) and MC (red) vs run date.
    \label{fig:angres} }
\end{figure}

Figure \ref{fig:angres} shows the distribution of the cosine of the
angle between the fit and `generated' direction for data and MC \ns~
events for a run taken at the centre of the detector.  In order to
characterize the these distributions, the following simple three
parameter function
\begin{equation}
F(\cos{\theta}) = N[e^{P_{2}(1-\cos{\theta})} + P_{1}e^{P_{3}(1-\cos{\theta})}]
\end{equation}
is fit to the distributions.  The overall normalization is arbitrary
but left free in the fit.  As can be discerned in Figure
\ref{fig:angres}, there is good agreement between the data and MC fit
parameters for this run.

Figure \ref{fig:angres} shows distributions of the three fit
parameters for data and MC runs as a function of run date.  Systematic
uncertainties are derived from these distributions are dominated by
the mean difference between the data and MC.  Analogous distributions
have been evaluated as a function of radial position within the D$_2$O
volume with the total systematic uncertainty on these parameters
assessed to be at the 10-15\% level.  This translates into negligible
uncertainties on the CC and NC measurements but does contribute
substantially to the ES total systematic uncertainty.  The ES
measurement is, however, still dominated by statistical uncertainties.

The most significant change with the inclusion of salt corresponds to
the change in response to NC events (neutrons).  In this
configuration, neutrons are primarily captured on $^{35}$Cl with a
subsequent release of multiple  photons
\begin{equation}
n + {^{35}Cl} \rightarrow  {^{36}Cl} + \Sigma\gamma \;\;\;(E_{\Sigma\gamma}\simeq
8.6 MeV).
\end{equation}

Thus not only does the higher capture cross-section on Cl dramatically
improve the neutron detection efficiency, but in addition the the
multiple photon nature of these events causes them to exhibit a
greater degree of isotropy than, for example, CC events with a single
Cerenkov cone.

\begin{figure}[htbp]
  \centerline{\hbox{ \hspace{0.2cm}
    \includegraphics[width=7.5cm]{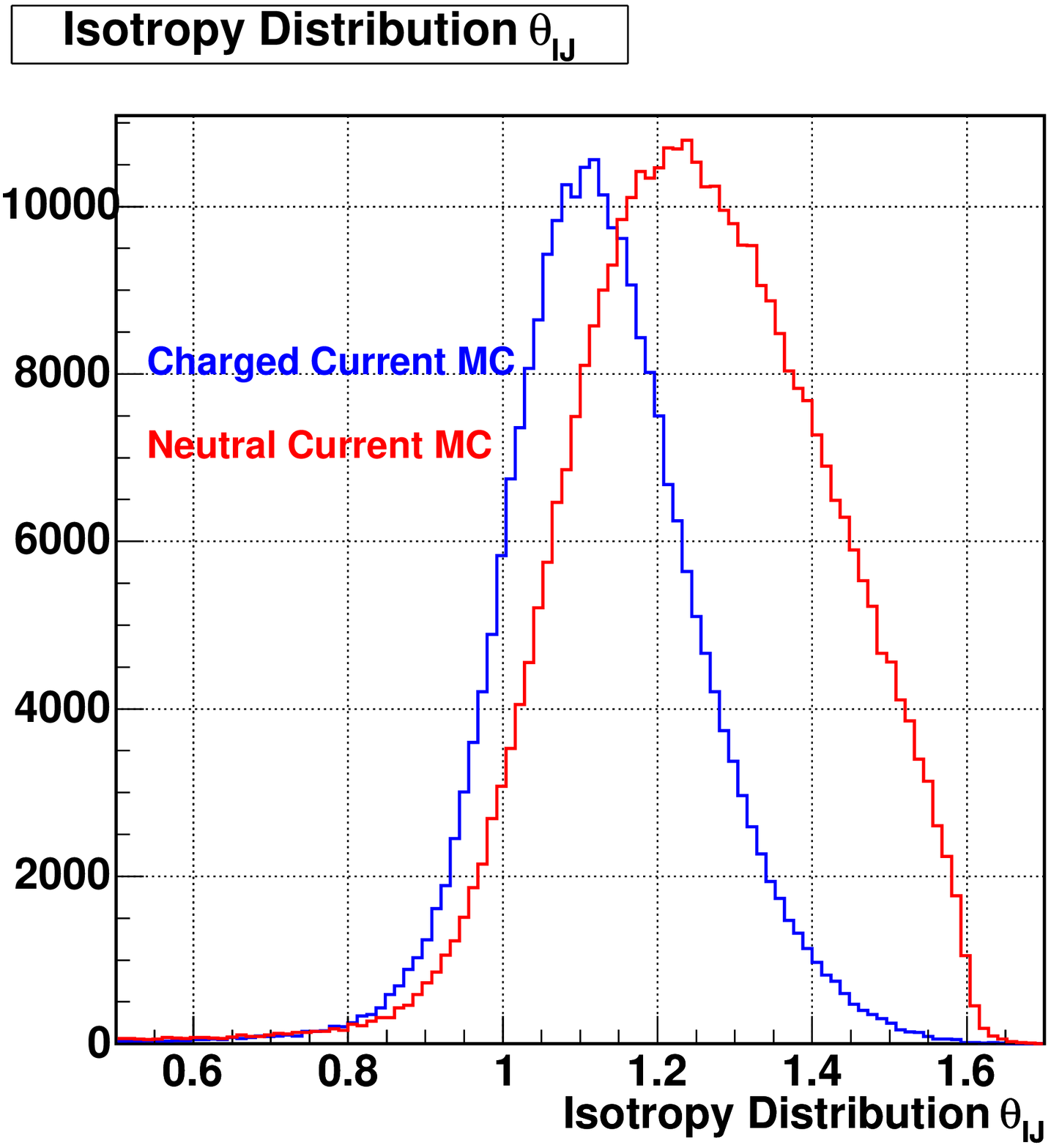}
    \hspace{0.15cm}
    \includegraphics[width=7.5cm]{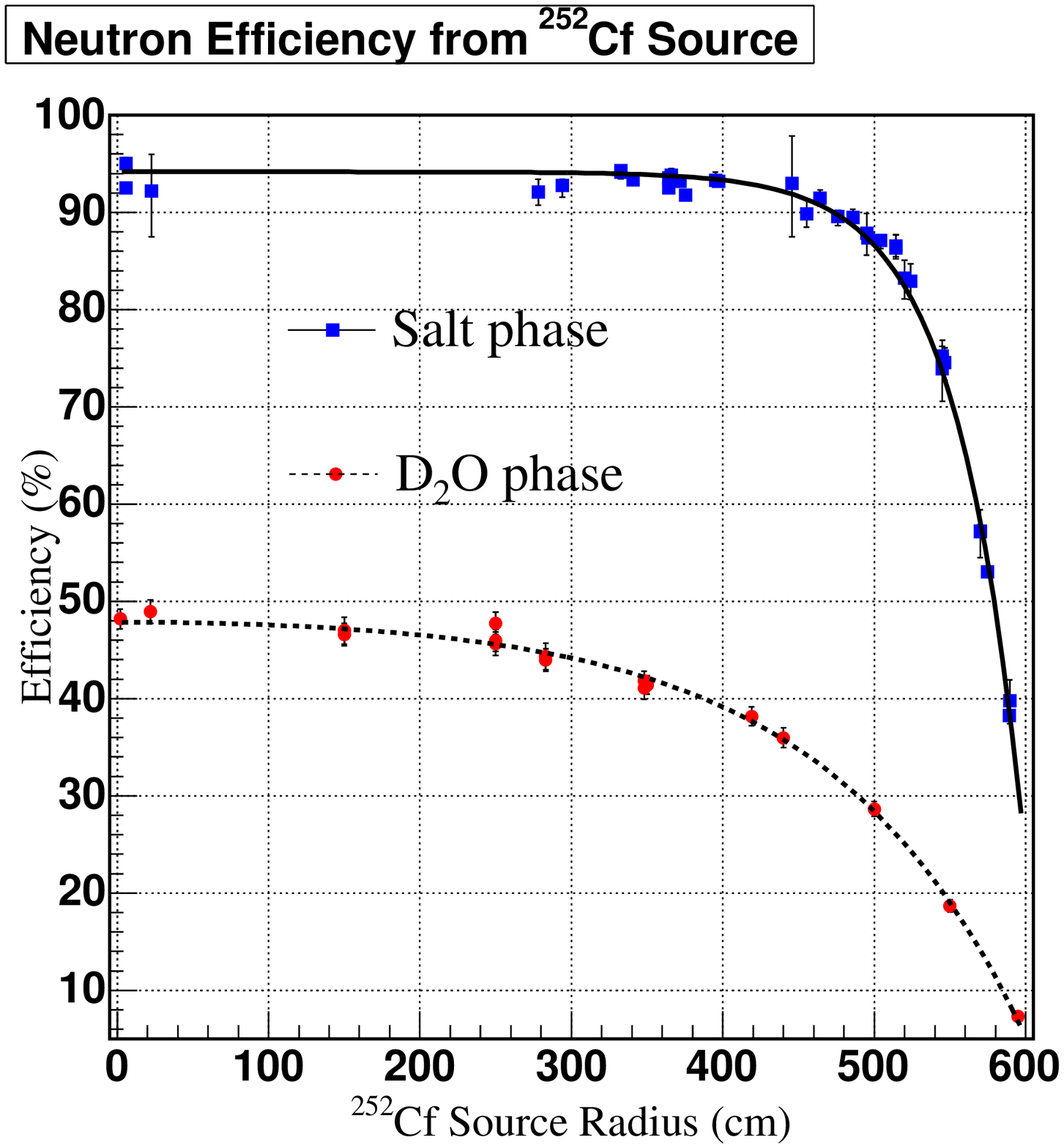}
    }
  }
 \caption{\it
      Shown are comparisons between CC (blue) and NC (red) MC  $\theta_{ij}$
      isotropy distributions (left) and between pure D$_2$O (red) and
      salt (blue) phase neutron detection efficiency distributions vs
      radial position (right).
    \label{fig:nresponse} }
\end{figure}

Figure \ref{fig:nresponse} compares CC and NC MC
simulated events for isotropy parameter $\theta_{ij}$.  This is the mean pair
angle between PMT's that have triggered in an event.  As can be seen,
such information is clearly useful in separating CC and NC events.
In fact, utilizing this information will allow a precise determination
of the fluxes to be made without the use of the energy spectrum.

The neutron efficiency has been measured via analysis of neutrons
produced from a $^{252}$Cf source placed at various positions within
the D$_2$O volume.  Figure \ref{fig:nresponse} shows the comparison
of neutron detection efficiencies between the pure D$_2$O and salt
phases as a function of radial position within the D$_2$O.  The
response at the centre of the detector is almost a factor of 
2 higher in salt but, owing to the greater difference at high radii,
the detection efficiency inside R=550 cm is improved by almost a factor of 3.

\subsection{Signal Extraction}

Signal extraction for the salt phase is carried out in the same
fashion as pure D$_2$O phase.  The neutrino candidate event
distributions are fit, using an unbinned extended maximum likelihood
technique, with NC, CC, and ES signal pdf's plus with fixed background
pdf's included.  The primary difference for the salt phase is the
altered shape of the NC pdf's and the inclusion of an additional
isotropy variable.

In order to evaluate the expected statistical sensitivities of the salt
phase dataset, signal MC simulated samples with rates corresponding to the
measured pure D$_2$O fluxes were generated fit with various
combinations of salt pdf's.  One live year of data was simulated and
cuts corresponding to the pure D$_2$O selection were applied.

Table \ref{tab:mcresults} presents the
percent statistical error from the published pure D$_2$O results and
for the salt MC study.  When all four variables (energy, radius, angle
between fit direction and direction to sun, and isotropy) are included
in the fit, the CC and ES uncertainties are projected to be similar to
the pure D$_2$O but the NC measurement improves by almost a factor of
2.  
\begin{table}[h]
\centering
\caption{ \it Statistical uncertainty MC projections for salt phase.  Variables are E=energy,
  R=event radius, $\theta_{sun}$=angle between fit direction and
  direction to sun, and $\beta_{14}$=isotropy parameter.
}
\vskip 0.1 in
\begin{tabular}{|c|c|c|c|c|} \hline
Phase &  Variables in Fit         & CC Error & NC Error & ES Error \\
\hline
\hline
Pure D$_2$O & E,R,$\theta_{sun}$              & 3.4\% & 8.6\% & 10\%\\
Salt MC     & E,R,$\theta_{sun}$              & 4.2\% & 6.3\% & 10\%\\
Salt MC     & E,R,$\theta_{sun}$,$\beta_{14}$ & 3.3\% & 4.6\% & 10\%\\
Salt MC     & R,$\theta_{sun}$,$\beta_{14}$  & 3.8\% & 5.3\% & 10\%\\
\hline
\end{tabular}
\label{tab:mcresults}
\end{table}

Interestingly, when the energy pdf's are not included in the fit
the CC uncertainty does not increase appreciably and the NC
uncertainty remains substantially better than in the pure D$_2$O
phase.  Hence with the salt phase data a precision measurement will be
made with no assumptions on the shapes of the signal energy distributions.

\section{Summary}

Precise measurements of the CC and NC fluxes from the pure D$_2$O
phase of SNO running have been presented.  These results can be translated
into values for electron neutrino and non-electron neutrino fluxes.
The measurement of a statistically significant non-electron neutrino flux
provides strong evidence for flavour transformation of electron
neutrinos produced in the sun and can be consistent with an MSW oscillation model.
The $^8$B shape constrained measurement of the total $^8$B neutrino
flux is consistent with standard solar model predictions.   The
unconstrained measurement, while still consistent,  is dominated by
statistical uncertainties. 

Various aspects of detector calibration and response for the salt
phase have been presented including energy scale, angular resolution, and
neutron response.  As indicated, the salt phase detector response is
well understood and these analyses are in an advanced stage in
preparation for the forthcoming initial salt phase publication.  

MC studies indicate that 1 year of salt livetime will produce a CC
result with similar statistical precision as the pure D$_2$O phase and
an NC result with substantially improved statistical precision.  Note
that the currently published NC measurement uncertainty is dominated
by statistics.  In addition, it has been demonstrated that a precision
energy-unconstrained fit can be made with the salt phase data.

\end{document}